\documentclass[11pt]{article}
\usepackage{epsfig}

\let\a=\alpha \let\b=\beta   
    
  \let\n=\nu

\def\nn{\nonumber} \def\bd{\begin{document}} \def\ed{\end{document}}
\def\ds{\documentstyle} \let\fr=\frac \let\bl=\bigl \let\br=\bigr
\let\Br=\Bigr \let\Bl=\Bigl 
\let\bm=\bibitem
\let\na=\nabla
\let\pa=\partial \let\ov=\overline 
\newcommand{\be}{\begin{equation}} 
\newcommand{\ee}{\end{equation}} 
\def\ba{\begin{array}}
\def\ea{\end{array}}
\def\ft#1#2{{\textstyle{{\scriptstyle #1}\over {\scriptstyle #2}}}}
\def\fft#1#2{{#1 \over #2}}
\def\del{\partial}
\def\vp{\varphi}
\def\sst#1{{\scriptscriptstyle #1}}
\def\oneone{\rlap 1\mkern4mu{\rm l}}
\def\td{\tilde}
\def\wtd{\widetilde}
\def\ie{\rm i.e.\ }
\def\dalemb#1#2{{\vbox{\hrule height .#2pt
        \hbox{\vrule width.#2pt height#1pt \kern#1pt
                \vrule width.#2pt}
        \hrule height.#2pt}}}
\def\square{\mathord{\dalemb{6.8}{7}\hbox{\hskip1pt}}}
\newcommand{\ho}[1]{$\, ^{#1}$}
\newcommand{\hoch}[1]{$\, ^{#1}$}
\newcommand{\bea}{\begin{eqnarray}} 
\newcommand{\eea}{\end{eqnarray}} 
\newcommand{\ra}{\rightarrow}
\newcommand{\lra}{\longrightarrow}
\newcommand{\Lra}{\Leftrightarrow}
\newcommand{\ap}{\alpha^\prime}
\newcommand{\bp}{\tilde \beta^\prime}
\newcommand{\tr}{{\rm tr} }
\newcommand{\Tr}{{\rm Tr} } 
\def\0{{\sst{(0)}}}
\def\1{{\sst{(1)}}}
\def\2{{\sst{(2)}}}
\def\3{{\sst{(3)}}}
\def\4{{\sst{(4)}}}
\def\5{{\sst{(5)}}}
\def\6{{\sst{(6)}}}
\def\7{{\sst{(7)}}}
\def\8{{\sst{(8)}}}
\def\n{{\sst{(n)}}}
\def\tV{\widetilde V}
\def\tW{\widetilde W}
\def\tH{\widetilde H}
\def\tE{\widetilde E}
\def\tF{\widetilde F}
\def\tA{\widetilde A}
\def\im{{{\rm i}}}
\def\tY{{{\wtd Y}}}
\def\ep{{\epsilon}}
\def\vep{{\varepsilon}}
\def\R{\rlap{\rm I}\mkern3mu{\rm R}}
\def\cD{{\cal D}}
\def\semi{{\ltimes}}

\newcommand{\NP}{Nucl. Phys. }
\newcommand{\tamphys}{\it Center for Theoretical Physics,
Texas A\&M University, College Station, TX 77843}
\newcommand{\upenn}{\it Dept. of Phys. and Astro.,
University of Pennsylvania,
Philadelphia, PA 19104}

\newcommand{\auth}{M. Cveti\v{c}\hoch{\dagger1},
H. L\"u\hoch{\dagger1} and C.N. Pope\hoch{\ddagger2}}

\begin{document}
\begin{flushright}
\hfill{CTP TAMU-11/00 \\
UPR/884-T \\
March 2000}\\
\hfill{\bf hep-th/0004201}\\
\end{flushright}

\vspace{10pt}

\begin{center}
{\large {\bf Consistent Sphere Reductions and Universality of 
the Coulomb Branch in the Domain-Wall/QFT Correspondence}}

\vspace{20pt}

\auth

\vspace{10pt}
{\hoch{\dagger}\upenn}

\vspace{10pt}
{\hoch{\ddagger}\tamphys}

\vspace{30pt}

\underline{ABSTRACT}
\end{center}

     We prove that any $D$-dimensional theory comprising gravity, an
antisymmetric $n$-index field strength and a dilaton can be
consistently reduced on $S^n$ in a truncation in which just $n$ scalar
fields and the metric are retained in $(D-n)$-dimensions, provided
only that the strength of the couping of the dilaton to the field
strength is appropriately chosen.  A consistent reduction can then be
performed for $n\le 5$; with $D$ being arbitrary when $n\le 3$, whilst
$D\le 11$ for $n=4$ and $D\le 10$ for $n=5$.  (Or, by Hodge
dualisation, $n$ can be replaced by $(D-n)$ in these conditions.)  We
obtain the lower dimensional scalar potentials and construct
associated domain wall solutions.  We use the consistent reduction
Ansatz to lift domain-wall solutions in the $(D-n)$-dimensional theory
back to $D$ dimensions, where we show that they become certain
continuous distributions of $(D-n-2)$-branes.  We also examine the
spectrum for a minimally-coupled scalar field in the domain-wall
background, showing that it has a universal structure characterised
completely by the dimension $n$ of the compactifying sphere.

{\vfill\leftline{}\vfill
\vskip 10pt \footnoterule {\footnotesize \hoch{1} Research supported
in part by DOE grant DOE-FG02-95ER40893
\vskip  -12pt} \vskip   14pt
{\footnotesize
        \hoch{2}        Research supported in part by DOE
grant DOE-FG03-95ER40917 \vskip -12pt}  \vskip  14pt
}

\pagebreak
\setcounter{page}{1}

\section{Introduction}

     The ability to embed a lower-dimensional theory in a
higher-dimensional one has proved to be an extremely useful one in
string theory.  One can, for example, re-interpret lower-dimensional
$p$-brane solitons as solutions of the ten-dimensional string, or
eleven-dimensional M-theory.  A crucial aspect of this picture is that
the Kaluza-Klein reduction must be a {\it consistent} one, in the
sense that all solutions of the lower-dimensional theory must also be
solutions of the original higher-dimensional theory.  This consistency
is guaranteed in a standard toroidal reduction, but it is far less
clear-cut when a reduction on a manifold such as a sphere is
considered.

    Kaluza-Klein reductions on spheres are of great interest in the
framework of string theory, because they can give rise to
lower-dimensional gauged supergravities that are relevant for
discussing the AdS/CFT correspondence \cite{malda,gkp,wit}.  The
generic structure of these gauged supergravities comprises gravity
coupled to a set of Yang-Mills gauge fields, and a set of scalar
fields with a non-trivial potential, together, possibly, with
additional antisymmetric tensor fields.  A particular class of
solution that can be studied is extremal domain walls, which can be
viewed as charged black holes or black $p$-branes in the gauged
theory, in the extremal limit for which the electric or magnetic
charges actually vanish.  Thus these solutions are supported entirely
by the metric and certain scalar fields within the gauged
supergravity.

    It therefore becomes of interest to study the circumstances under
which a higher-dimensional theory can admit a consistent $n$-sphere
reduction in which just gravity and appropriate scalar fields are
retained.  In some cases this may be viewed as a subset of a larger
consistent reduction of a gauged supergravity, in which the starting
point is supergravity in ten or eleven dimensions.  However, the
question can also be posed in a more general framework, where the
starting point need not necessarily even be a supersymmetric theory.

    Before discussing the possible new cases, let us review what is
known at present.  It is natural, when considering an $n$-sphere
reduction, to try to retain all the $SO(n+1)$ Yang-Mills fields as
part of the consistent reduction.  Usually, however, this is not
possible.  It was recently shown in \cite{clps3,clpst} that the cases
where this can be done, starting from a $D$-dimensional theory of
gravity, $n$-form field strength and dilaton, are as follows.  One can
start with $(D,n)=(11,4)$, and reduce on $S^4$ or $S^7$ to seven or
four dimensions respectively; another possibility is to start from
$(D,n)=(10,5)$, and reduce on $S^5$ to five dimensions.  In these
cases, the system has no dilaton.  Including a dilaton, with a
specific coupling, one can also start with $n=3$ and $D$ arbitrary,
reducing on $S^3$ or $S^{D-3}$; or finally one can start with $n=2$
and $D$ arbitrary, and reduce on $S^2$.  In all cases one must also
include scalar fields $T_{ij}$ in the reduction Ansatz, corresponding
to the coset $SL(n+1,\R)/SO(n+1)$.  Additionally, for the $S^4$
reduction of $D=11$ one must include five 3-form field strengths in
the Ansatz, while for the $S^7$ reduction one must also include 35
more pseudoscalars.\footnote{In fact the consistent reductions from
$D=11$ require, in addition, the inclusion of the $FFA$ in the
Lagrangian that arises in $D=11$ supergravity.  In the $S^5$ reduction
from $D=10$, it is necessary to impose the requirement of self-duality
on the 5-form field strength.}  Finally, for the $S^3$ reduction in
the $n=3$ case, one must include a 3-form field strength in the
reduction Ansatz.

    Our statement of the possible consistent reductions first
specified that the $SO(n+1)$ Yang-Mills gauge fields were to be
included, and then we listed the additional fields that would be
needed for consistency.  Another way of phrasing the question is to
specify which scalar fields will be included in the reduction Ansatz.
In fact if we want to include all the scalars $T_{ij}$ of the
$SL(n+1,\R)/SO(n+1)$ coset, the list of cases where consistent
reductions are possible will be the same as the above.  The reason for
this is that once all the scalars $T_{ij}$ are present, they will act
as sources for the Yang-Mills gauge fields, and so it would be
inconsistent to omit the Yang-Mills fields.  However, if we settle for
a reduction in which fewer scalars are retained, it becomes possible
to omit the Yang-Mills fields and this opens up some further
possibilities for consistent reductions, which we shall explore in
this paper.  These reductions with scalars but no gauge fields will be
sufficient for the purpose of constructing the extremal domain-wall
solutions in the lower dimension, and then lifting them back to the
higher dimension.

   As mentioned above, if one includes the full set of scalars
$T_{ij}$ in a truncation then they will give rise to source terms that
require the Yang-Mills fields to be non-zero.  Specifically, the 
source currents are of the form $T_{k[i}^{-1}\, \del_\mu \, T_{j]k}$,
in the adjoint of $SO(n+1)$.  If we make a truncation where only the
diagonal scalar fields are retained,
\be
T_{ij} = {\rm diag}(X_1,X_2,\ldots X_{n+1})\,,\label{diag}
\ee
then the currents $T_{k[i}^{-1}\, \del_\mu \, T_{j]k}$ will be zero,
and thus there is no longer any necessity to include the gauge fields
in a consistent truncation.   This actually allows a somewhat 
extended set of $(D,n)$ values for which consistent reductions can be
achieved, which includes cases that would not allow consistent
reductions with $SO(n+1)$ gauge fields.  The allowed cases are
detailed below.

    In section 2 we construct an Ansatz for the $n$-sphere reduction of the a
$D$-dimensional theory of gravity, an $n$-form field strength, and a
dilaton, in which the lower-dimensional fields comprise just gravity
and the diagonal scalar fields given by (\ref{diag}).  We obtain a
complete proof of the consistency of this Kaluza-Klein reduction,
showing that it works in all cases where the strength of the coupling
of the dilaton to the $n$-form in $D$ dimensions is appropriate.  This
requirement on the coupling is a rather stringent one, and the
allowable cases turn out to be $\{n=5, D\le 10\}$; $\{n=4, D\le 11\}$; 
and $n\le3$ with $D$ arbitrary, for $n\le D/2$.

    In section 3 we construct $(n+1)$-parameter extremal domain-wall
solutions in the lower-dimensional theories of gravity plus scalar
fields, and then make use of the reduction Ansatz derived in section 2
in order to lift these solutions back to the original $D$-dimensional
theory.  We show that in the higher dimension the lifted solutions
admit an interpretation as continuous distributions of
$(D-n-2)$-branes.  We discuss and obtain the distribution functions.
We obtain the metric of the distributed branes in the dual frame, and
show that the structure of these metrics depends only on the dimension
$n$ of the internal sphere, but is independent of $D$.  In particular,
the metric in the dual frame becomes asymptotically AdS$\times S^n$
for $n\ne 3$, and Minkowski$\times S^3$ for $n=3$.

   In section 4 we analyse the spectrum of excitations of a
minimally-coupled scalar in the background of the $(D-n)$-dimensional
domain-wall solution, showing that it has a universal structure that
is characterised by the dimension $n$ of the internal sphere used in
the dimensional reduction.  In the case of the vacuum solutions, where
the $(n+1)$ parameters in the general solutions are all set to zero,
the scalar wave equation can be solved explicitly, allowing a study of
the singularity structure.  We also analyse the Schr\"odinger
potentials for generic cases, allowing us to determine the
structures of the spectra in the various cases.

       In an appendix, we show that a single-charge rotating $p$-brane in
a generic dimension can be dimensionally reduced on the internal
(distorted) $n$-sphere to give rise to domain-wall black holes with
$[(n+1)/2]$ electric $U(1)$ charges.   In the extremal limit, the
gauge fields vanish and the balck hole becomes a domain wall that is
contained within the set of solutions obtained in this paper.

\section{Kaluza-Klein sphere reduction}

      Single-charge $p$-branes in supergravity theories in $D$
dimensions can be classified as solutions of the theory described by
the Einstein-Hilbert action
coupled to a dilaton and an $n$-form field strength,
\be
{\cal L}_D = \hat R\, {\hat*\oneone} - 
\ft12{\hat *d\hat\phi}\wedge d\hat\phi -\ft12\, e^{a\hat\phi}\,
{\hat *\hat F_\n}\wedge \hat F_\n\,,\label{genlag}
\ee
where the constant $a$ is given by \cite{dkl}
\be
a^2 = 4 - \fft{2(n-1)(D-n-1)}{D-2}\,.\label{avalue}
\ee
The requirement that $a$ be real puts a strong condition on the
possible values for $n$, bearing in mind that we must have $n\le D$,
and in fact we can always choose a dualisation for $F_\n$ for which
$n\le D/2$.  From (\ref{avalue}), it then follows that the 
maximum value for $n \le D/2$ is 5.  Whe $n=5$,
the maximal dimension is $D=10$, corresponding to the self-dual 5-form
in the type IIB theory.  For $n=4$, the maximal dimension is $D=11$,
corresponding to 11-dimensional supergravity.  In both cases, the
constant $a$ vanishes.  For $n=0,1,2,3$, the dimension $D$ can be
arbitrary.  Note that for a given $n$ satisfying ({\ref{avalue}),
$n'=D-n$ satisfies it too.  To summarise, the allowed possibilities are
\bea
n=5\,,\quad D-5:&& D\le 10\,\nn\\
n=4\,,\quad D-4:&& D\ge 11\,\nn\\
n=0,1,2,3\,,\quad D,D-1,D-2,D-3:&& D\quad \hbox{arbitrary}\,.
\label{possible}
\eea
Note that these results come from the requirement only that $a$ must be
real.  If in addition we require that the Lagrangian must be associated
with a supersymmetric theory, we get the further restriction that the
dimension $D$ must be less than or equal to eleven or ten.

        The $p$-branes for which the first term on the right-hand-side
of (\ref{avalue}) is 4 can be viewed as the basic building blocks for
$p$-brane solitons.  The $p$-branes with values other than 4, (usually
$4/N$ with $N$ an integer) can be viewed as bound states or
intersections of these building blocks.  For example, for $D=11$ and
$D=10$, our discussion applies to M-branes, the NS-NS string and
5-brane and all the D-branes.

         We shall now consider the Kaluza-Klein dimensional reduction
of the Lagrangian (\ref{genlag}) on $S^n$.  (The discussion of the
reduction instead on $S^{D-n}$ can be handled by dualising the $n$-form field
strength to a $(D-n)$-form.)  In general, such a reduction is
inconsistent if we keep all the massless fields.  It was shown
\cite{clps3}, however, that for $n=2$ and $n=3$ the reduction is always
consistent, provided that (\ref{avalue}) is satisfied.  For $n=5$ and
$n=4$, the reduction is consistent only if additional conditions are
satisfied, namely self-duality of the 5-form in $D=10$, and the
addition of an $FFA$ term in $D=11$ for the $n=4$ case.

        In this paper, we shall truncate further to a subset of the
massless fields, corresponding to ``diagonal inhomogeneous
distortions'' of the internal $S^n$ metric.  By this, we mean that we
canonically embed the sphere $S^n$ in $n+1$ dimensional Euclidean
space.  The round $S^n$ metric is given by $d\mu_i d\mu_i$, where
$\mu_i$ are Euclidean coordinates satisfying the unit-length
constraint $\mu_i\, \mu_i =1$.  The diagonal inhomogeneous distortion of
the sphere is then achieved by introducing $(n+1)$ scalars $X_i$, and scaling
the coordinate differentials as follows:
\be
ds^2_n= \sum_i X_i^{-1} (d\mu_i)^2\,.\label{snmetric}
\ee 
We shall show that for this subset of fields, the Kaluza-Klein 
reduction is consistent for any of the $D$ and $n$ values listed in
(\ref{possible}), provided that (\ref{avalue}) is satisfied. 

     We find that the Kaluza-Klein reduction Ansatz is given by
\bea
&&d\hat s_D^2 = Y^{\fft{1}{D-2}}\Big(\Delta^{\fft{n-1}{D-2}}\, ds_{D-n}^2 +
g^{-2}\,\Delta^{-\fft{(D-n-1)}{(D-2)}}\sum_{i=1}^{n+1} X_i^{-1} 
(d\mu_i)^2 \Big)
\,,\nn\\
&&e^{-\fft{2}{a}\hat\phi} = \Delta^{-1}\, Y^{\fft{2(D-n-1)}{a^2(D-2)}}
\,,\label{ansatz}\\
&&\hat F_\n = g^{-n+1}\, \Delta^{-2}\, U\, W + g^{-n+1}\,
\del_\nu\Big( \fft{X_i\, \mu^i}{\Delta}\Big)\, dx^\nu \wedge
 Z_i\,.\nn
\eea
where
\be
\mu_i \mu_i = 1\,,\quad Y = \prod X_i\,,\quad
\Delta= \sum X_i\, \mu_i^2\,,\quad
U = 2\sum_i X_i^2\, \mu_i^2 - \Delta\, \sum_i X_i\,.
\ee
The quantities $W$ and $Z_i$ are respectively the volume-form on the
$n$-sphere, and a certain $(n-1)$-form on the $n$-sphere:
\bea 
&&W=\fft1{n!}\, \ep_{i j_1\cdots j_n }\mu^i\, d\mu^{j_1}\wedge \cdots
\wedge d\mu^{j_n}\,,\\
&&Z_i = \fft1{(n-1)!}\, \ep_{ij_1\cdots j_n}\, \mu^{j_1}\,
d\mu^{j_2}\wedge \cdots \wedge d\mu^{j_n}\,.\nn
\eea
We find after some algebra that the dual of the field strength $F_\n$
is given by
\be
e^{a\hat \phi}\, {\hat *F_\n} = g\, U\, \epsilon_{D-n} +
\fft1{2g}\, X_i^{-1} {*dX_i}\wedge d(\mu_i^2)\,.
\ee

     We can then substitute the Ansatz into higher dimensional
equations of motion. First, we can verify that the Ansatz for $\hat
F_\n$ in (\ref{ansatz}) satisfies the Bianchi identity $d\hat
F_\n=0$.  Next, we look at the equations of motion for the field
strength $\hat F_\n$ and the dilaton $\hat\phi$:
\bea
d\Big( e^{a\hat\phi}\, {\hat *\hat F_\n}\Big) &=& 0\,,\nn\\
(-1)^D\, d{\hat*d\hat\phi} &=& -a\, e^{a\hat\phi}\, {\hat *\hat
F_n}\wedge \hat F_\n\,.
\eea
After a considerable amount of algebra, we find that the Ansatz yields
a consistent dimensional reduction of these
$D$-dimensional equations to give the following $(D-n)$-dimensional equations
for the scalar fields:
\bea
&&(-1)^{D-n}\, d(\wtd X_i^{-1}\, {* d\wtd X_i}) = - 2g^2\,Y^{\fft2{n+1}} \,
\Big[2 \wtd X_i^2 -
\wtd X_i\, \sum_j \wtd X_j -\ft{2}{n+1} \sum_j \wtd X_j^2 
+ \ft1{n+1}\, (\sum_j
\wtd X_j)^2 \Big]\, \ep_{D-n}\,,\nn\\
&&(-1)^{D-n}\, \ft{2(D-n-2)}{(D-2)\, a^2}\, d(Y^{-1}\, {*dY}) =
-V\, \ep_{D-n}\,.
\eea
Here, we have defined the rescaled fields $\wtd X_i$ by
\be
X_i = Y^{\fft1{n+1}}\, \wtd X_i\,,
\ee
so that $\prod_i \wtd X_i=1$, and the potential $V$ is defined by
\be
V\equiv  \ft12 g^2\, \Big(2 \sum_i X_i^2 - (\sum_i X_i)^2 \Big) = 
\ft12 g^2\, Y^{\fft2{n+1}}\, 
\Big(2 \sum_i \wtd X_i^2 - (\sum_i \wtd X_i)^2 \Big)\,.
\ee

    Finally, to check the higher-dimensional Einstein equations, we need first
to calculate the Ricci tensor for the metric in (\ref{ansatz}).  This is most
easily done by noting that it is conformally related to the metric
\be
d\bar s^2_D = \Delta^{p}\, ds_{D-n}^2 + \Delta^{-q}\, \sum_i
X_i^{-1}\, (d\mu_i)^2\,,\label{barmetric}
\ee
with 
\be
d\hat s_{D}^2 = e^{2f}\, d\bar s^2_D \,,
\ee
where we have defined
\be
e^{2f}= Y^{\fft1{D-2}}\,,\qquad p= \fft{n-1}{D-2}\,,\qquad q=
\fft{D-n-1}{D-2}\,.
\ee
It is easy to establish the standard result that the coordinate-frame
components of the Ricci tensor $\hat R_{MN}$ for the metric $d\hat
s_D^2$ are related to the coordinate-frame components $\bar R_{MN}$
for the metric $d\bar s_{D}^2$ by
\be
\hat R_{MN} = \bar R_{MN} +
(D-2)\Big(\del_M f\, \del_N  f -\bar\nabla_M\, \del_N\, f -\bar
g^{PQ}\, (\del_P f)(\del_Q   f)\, \bar g_{MN}\Big)-
\bar{\square} f\, \bar g_{MN} \,.\label{conformalmet}
\ee
Results for the Ricci tensor for certain metrics of the form 
(\ref{barmetric}) were derived in \cite{clps}, and with minor
modifications they can be carried over to our present case.
They were obtained in a basis where one of the $(n+1)$ coordinates $\mu_i$,
say $\mu_0$, is expressed in terms of the $n$ remaining ones $\mu_\a$
by using the relation $\mu_i\, \mu_i=1$.  Thus the components
$g_{\a\b}$ of the distorted $n$-sphere metric (\ref{snmetric}), and
its inverse, are given by
\bea
g_{\a\b} &=& X_\a\, \delta_{\a\b} + X_0^{-1}\, \hat \mu_\a\, \hat
\mu_\b\,,\nn\\
g^{\a\b} &=& X_\a\, \delta_{\a\b} - \Delta^{-1}\, X_\a\, X_\b\,
\mu_\a\, \mu_b\,,
\eea
where in the first line we are writing $\hat \mu_\a = \mu_\a/\mu_0$.
We refer to \cite{clps} for many of the details of the curvature
calculations.  Combining these results with (\ref{conformalmet}), we
obtain, after extensive algebraic manipulations, the following
expressions for the lower-dimensional spacetime, internal and mixed
components of the $D$-dimensional Ricci tensor:
\bea
\hat R_{\mu\nu} &=& R_{\mu\nu} -\ft{(n-1)(D-n-1)}{4(D-2)}\,
\Delta^{-2}\, \del_\mu\Delta\, \del_\nu\Delta - \ft12p\, \Delta^{-1}\,
\square\Delta\, g_{\mu\nu} + \ft12 p\, \Delta^{-2}\,
\del_\lambda\Delta\, \del^\lambda\Delta\, g_{\mu\nu}\nn\\
&&-\ft14 X_i^{-2}\, \del_\mu X_i\, \del_\nu X_i + \ft12 \Delta^{-1}\,
X_i^{-1}\, \mu_i^2\, \del_\mu X_i\, \del_\nu X_i + \ft1{4(D-2)}\,
Y^{-2}\, \del_\mu Y\, \del_\nu Y \nn\\
&&- \ft14 q\, \Delta^{-1}\, (\del_\mu
\Delta\, \del_\nu Y + \del_\nu\Delta\, \del_\mu Y) 
 - \ft1{2(D-2)}\, \square \log Y\, g_{\mu\nu} \nn\\
&&- p\, \Big( \sum_i X_i^2 - \Delta^{-1}\, X_i^2\, \mu_i^2\, \sum_j X_j
  + 2\Delta^{-2}\, (X_i^2\, \mu_i^2)^2 - 2\Delta^{-1}\, X_i^3\,
  \mu_i^2\Big)\, g_{\mu\nu}\,,\nn\\
\hat R_{\a\beta} &=& R_{\a\beta} +\ft12 q\, 
g_{\a\beta}\, \Delta^{-2}\, 
\square\, \Delta - \ft12 q\,  g_{\a\beta}\,\Delta^{-3}\,  
\del_\lambda\Delta\, \del^\lambda\Delta-\ft12 \Delta^{-1}\, \square\,
g_{\a\beta} \label{riccires}\\
&&+ \ft12 \Delta^{-1}\, g^{\gamma\delta}\, \del_\lambda
g_{\a\gamma}\, \del^\lambda g_{\beta\delta}
 -\ft14 \Delta^{-2}\, \del_\a\Delta\, \del_\beta\Delta - \ft12
\Delta^{-1}\, \nabla_\a\del_\beta\, \Delta \nn\\
&&- \ft14 q\, g_{\a\beta}\,\Delta^{-2}\,
\del_\gamma\Delta\, \del^\gamma \Delta+ \ft12 q\,
g_{\a\beta}\, \Delta^{-1}\, \nabla_\gamma\del^\gamma\, \Delta
-\ft{1}{2(D-2)}\, \Delta^{-1}\, \square\log Y\, g_{\a\beta} 
\,,\nn\\
\hat R_{\a\mu} &=& - \ft12 \Delta^{-2}\, U\, ( X_\a^{-1}\, \del_\mu
X_\a - X_0^{-1}\, \del_\mu X_0)\, \mu_\a + \ft18 a^2\, \Delta^{-2}\,
\del_\mu\Delta\, \del_a\Delta - \ft14 q\, \Delta^{-1}\, Y^{-1}\,
\del_\mu Y\, \del_\a\Delta\,.\nn
\eea
Note that here we are using a ``generalised'' summation convention in
which summations over the $i$ index, where not otherwise indicated,  
are understood.  The $\square$ operator denotes the d'Alembertian
calculated in the lower-dimensional metric $g_{\mu\nu}$, and
$R_{\a\beta}$ denotes the Ricci tensor of the internal metric (\it i.e.}\
the Ricci tensor for the metric (\ref{snmetric}), with the $X_i$ are
treated as parameters independent of the internal coordinates).  

    The $D$-dimensional Einstein equation reads $\hat R_{MN} = \hat
S_{MN}$, where
\be
\hat S_{MN}= \ft12 \del_M\hat\phi\, \del_N\hat\phi + \fft{e^{a\, \hat \phi}}{2(D
-n-1)!}\, \Big( \hat F_{MN}^2
- \fft{D-n-3}{(D-n)(D-n-1)}\, \hat F^2\, \hat g_{MN}\Big)\,.
\ee
After some algebra we find that $\hat S_{MN}$ is given by
\bea
\hat S_{\mu\nu} &=& \ft12 \Delta^{-1}\, \mu_i^2\, X_i^{-1}\, \del_\mu
X_i\, \del_\nu X_i - 
\ft{(n-1)(D-n-1)}{4(D-2)}\,  \Delta^{-2}\, \del_\mu\Delta\, \del_\nu\Delta 
+ \ft{(D-n-1)^2}{2 a^2\, (D-2)^2}\, Y^{-2}\, \del_\mu Y\, \del_\nu
Y\nn\\
&& - \ft14 q\, \Delta^{-1}\, Y^{-1}\, (\del_\mu\Delta\, \del_\nu Y +
\del_\nu\Delta\, \del_\mu Y)\nn\\
&& -\ft12 p\, \Delta^{-2}\, \Big( U^2 -
\del_\lambda\Delta\, \del^\lambda\Delta + \Delta \mu_i^2\, X_i^{-1}\,
\del_\lambda X_i\, \del^\lambda X_i\Big)\, g_{\mu\nu}\,,\nn\\
\hat S_{\a\beta} &=& \ft12 q\, \Delta^{-3}\, U^2\, g_{\a\beta} +
\ft12 q\,\Delta^{-2}\, g_{\a\beta}\, X_i^{-1}\, \mu_i^2\, \del_\lambda
X_i\, \del^\lambda X_i -\ft12 q\, \Delta^{-3}\, \del_\lambda \Delta\,
\del^\lambda \Delta\, g_{\a\b}\nn\\
&&-\ft12 \Delta^{-2}\,  (X_\a^{-1}\, \del_\lambda X_\a - X_0^{-1}\,
\del_\lambda X_0)(X_\beta^{-1}\, \del^\lambda X_\beta - X_0^{-1}\,
\del^\lambda X_0)\, \mu_\a\, \mu_\beta\nn\label{salbe}\\
&& + \ft18 a^2\, \Delta^{-2}\, \del_\a\Delta\, \del_\beta\Delta\,,\\
\hat S_{\a \mu} &=& -\ft12\Delta^{-2}\, U\, (X_\a^{-1}\, \del_\mu X_\a
- X_0^{-1}\, \del_\mu\, X_0)\, \mu_\a + \ft18 a^2\, \del_\mu\Delta\,
\del_\a\Delta - \ft14 q\, \Delta^{-1}\,  Y^{-1}\, \del_\mu Y\,
\del_\a\Delta\,.\nn
\eea

   After making use of the already-established equations of motion for
the scalar fields, we eventually find after considerable further
algebra that the $\hat R_{\mu\nu} = \hat S_{\mu\nu}$ components of the
higher-dimensional Einstein equation imply
\be
R_{\mu\nu} = \ft14 \wtd X_i^{-2}\, \del_\mu\wtd X_i\, \del_\nu\wtd X_i
+ \ft{2(D-n-2)}{(D-2)(n+1)\, a^2}\, Y^{-2}\, \del_\mu Y\, \del_\nu Y +
\ft1{D-n-2}\, V\, g_{\mu\nu}\,.
\ee
The full system of $(D-n)$-dimensional equations of motion can
therefore be derived from the Lagrangian
\be
{\cal L}= R\, {*\oneone}  - \ft{2(D-n-2)}{(n+1)(D-2)\, a^2}\,
Y^{-2}\, {*dY}\wedge dY  - \ft14 \sum_i \wtd X_i^{-2}\, {*d\wtd
X_i}\wedge d\wtd X_i - V\, {*\oneone}\,.\label{scalarlag}
\ee

   It remains to check the consistency of the other components of the
$D$-dimensional Einstein equations.  After making use of the
lower-dimensional equations of motion for the scalar fields, we find
that the internal components $\hat R_{\a\b}$ of the higher-dimensional
Ricci tensor agree precisely with the expression for $\hat
S_{\a\beta}$ that follows from substituting the Ans\"aze for $\hat
F_\n$ and $\hat\phi$, given in (\ref{ansatz}), into (\ref{salbe}).
Again, we have made extensive use of formulae derived in \cite{clps},
appropriately modified to the case under consideration here.  Finally,
we note that the mixed components $\hat R_{\a\mu}$ in (\ref{riccires})
agree precisely with the mixed components of $\hat S_{\a\mu}$ given in
(\ref{salbe}).

   With these calculations we have now obtained a complete and
explicit proof that the Ansatz (\ref{ansatz}) yields a consistent
Kaluza-Klein $n$-sphere reduction of the $D$-dimensional theory
described by (\ref{genlag}), with the lower-dimensional fields
appearing in the Ansatz satisfying the equations of motion that follow
from the $(D-n)$-dimensional Lagrangian (\ref{scalarlag}).

\section{Domain walls as distributions of $p$-branes}

     We find that the $d$-dimensional gravity/scalar Lagrangian
(\ref{scalarlag}) admits a domain wall solution, given by
\bea
ds_{d}^2&=& (gr)^{\fft{a^2(D-2)}{2(d-2)}}\Big((gr)^{n-3}\, 
h^{\ft1{2(d-2)}}\, dx^\mu dx_\mu + h^{-\ft{d-3}{2(d-2)}}\, 
\fft{dr^2}{g^2r^2}\Big)\,,\nn\\
X_i&=&(gr)^{\ft{a^2(D-2)}{4(d-2)}}\, h^{\ft{(d-3)}{4(d-2)}}\, H_i^{-1}
\,,\label{gendw}
\eea
where
\be
h\equiv \prod_{i=1}^{n+1} H_i\,,\qquad H_i = 1 +\fft{\ell_i^2}{r^2}
\,.
\ee

    In fact there is a redundancy in the paramtrisation of these
solutions, which can be seen as follows.  We make the following
transformation of the radial coordinate,
\be
r^2 = R^2 -L^2\,,\label{redo0}
\ee
where $L$ is a constant, and define new quantities as follows:
\be
\wtd H_i \equiv 1+\fft{\td\ell_i^2}{R^2}\,,\qquad \td h \equiv
\prod_{i=1}^{n+1}\wtd H_i\,,\qquad \td\ell_i^2 \equiv \ell_i^2-L^2\,.
\label{redo1}
\ee
After straightforward calculations, we find that the solution
(\ref{gendw}) becomes
\bea
ds_{d}^2&=& (gR)^{\fft{a^2(D-2)}{2(d-2)}}\Big((gR)^{n-3}\, 
\td h^{\ft1{2(d-2)}}\, dx^\mu dx_\mu + \td h^{-\ft{d-3}{2(d-2)}}\, 
\fft{dR^2}{g^2R^2}\Big)\,,\nn\\
X_i&=&(gR)^{\ft{a^2(D-2)}{4(d-2)}}\, \td h^{\ft{(d-3)}{4(d-2)}}\, \wtd
H_i^{-1} \,,\label{gendw2}
\eea
This is identical in form to the original solution (\ref{gendw}), but
with the redefined functions given in (\ref{redo1}).  Let us suppose
that, without loss of generality, the parameters $\ell_i$ are ordered
so that $\ell_1^2 \ge \ell_2^2\ge  \cdots \ge \ell_{n+1}^2$.  If we
choose the constant $L$ in the coordinate transformation (\ref{redo0})
to be equal to $\ell_{n+1}$, then we see that the original solution
with $(n+1)$ parameters $\ell_i$ (with $1\le i\le n+1$) is really
nothing but a solution with only $n$ parameters $\td\ell_i^2$ (with
$1\le i\le n$).   

          When $a=0$, which occurs for the cases $(D,n)=(11,4),
(11,7)$ and $(10,5)$, the resulting solutions become AdS domain walls.
The metrics in these cases become asymptotically-AdS spacetimes in
seven, four and five dimensions.  These AdS domain-wall solutions are
sphere reductions of the decoupling limits of ellipsoidal distributions
of M-branes and D3-branes.  These cases (and subsets) were studied
previously in \cite{KLT,FGPW,BS,BSI,dist,BBS}.

      In this paper, we shall extend the previous analysis to include
the cases where the dilaton-coupling constant $a$ is non-vanishing.
For these cases, the domain-wall metric (\ref{gendw}) is no longer
asymptotically AdS, but instead is asymptotic to a vacuum domain wall
as $r\rightarrow\infty$, given by
\be
ds_d^2 = \rho^{\fft{4(n+1)}{a^2(D-2)}}\, dx^\mu\, dx_\mu + 
g^{-2}\, d\rho^2\,.
\ee
where $\rho\sim (gr)^{\fft{a^2(D-2)}{4(d-2)}}$.  This metric is flat
as $\rho$ approaches at infinity.

       In the region near $r=0$, the metric structure depends on the
number of non-vanishing parameters $\ell_i$.  If $k$ of the $\ell_i$ 
are non-vanishing, we have
\be
ds_d^2 = \rho^{\gamma}\, dx^\mu dx_\mu + d\rho^2\,,\label{vacuum}
\ee
where
\be
\gamma = \fft{4(n+1-k)}{a^2(D-2) + 2(d-3)k}\,,\qquad
\rho = (gr)^{\fft{a^2(D-2) + 2(d-3)k}{4(d-2)}}\,.
\ee
Thus we see that at $r=0=\rho$, the solution is generic singular.  To
see if the singularity is naked or not, we evaluate
\be
\gamma -2 = \fft{4(d-2)(n-3-k)}{a^2(D-2)+2(d-3)k}\,.
\ee
Thus for $n=0,1,2,3$, the solution has a naked singularity for all
values of $k$.   For $n\ge 4$, the singularity is naked for $k>n-3$,
but marginal for $k\le n-3$.

        If we oxidise the solution back to $D$ dimensions, it acquires
an interpretation as a continuous distribution of $(D-n-2)$-branes,
given by
\bea
ds_D^2 &=& H^{-\ft{n-1}{D-2}}\, dx^\mu dx_\mu + H^{\fft{D-n-1}{D-2}}
\, dy^m dy^m\,,\nn\\
e^{-\fft{2}{a}\phi}&=& H\,.\qquad
F_\n = e^{-a\phi}{\hat *(d^{D-n-1}x\wedge dH^{-1})}\,.
\eea
where $H$ and the transverse Euclidean metric are given by
\bea
dy^m dy^m &=& h^{-\ft12}\, \wtd\Delta\, dr^2 + r^2\, \sum H_i\,
d\mu_i^2\,, \nn\\
H&=&\fft{1}{(gr)^{n-1}\, \wtd \Delta}\,,\qquad
\wtd \Delta = h^{\ft12}\sum\fft{\mu_i^2}{H_i}\,.
\eea
The function $H$ is a harmonic function of the Euclidean transverse
space, and it can be expressed as
\be
H=g^{-(n-1)}\, \int \fft{\sigma(\vec y{\,}')\, 
   d^{n+1}y'}{|\vec y -\vec y{\,}'|^{n-1}}\,,\label{hintegral}
\ee
where $\sigma (\vec y)$ is the distribution function.  The harmonic
functions in our cases here are associated with ellipsoidal
distributions.

      A detailed analysis is given in \cite{dist}, where
the charge-distribution functions are obtained in the non-dilatonic
cases of 3-branes in $D=10$, and M2-branes and M5-branes in $D=11$.
The analysis here is almost identical, and we shall not
enumerate all the possibilities.  It was observed in \cite{dist} that
although the results for the charge-distribution functions are
distinctly different depending upon how many of the $\ell_i$
parameters are non-zero, by carefully taking limits in which some of
the parameters are sent to zero one can view them  all as being derived
from a maximally-degenerate case with all $(n+1)$ parameters
non-zero.  The distribution function with all the $\ell_i$ non-vanishing
is given by \cite{dist}
\be
\sigma_{n+1} =\fft{1}{V_n\, \prod_{i=1}^{n+1} \ell_i}\, 
\delta'(1 - \sum_{i=1}^{n+1}
\fft{y_i^2}{\ell_i^2})\,,\label{np1charge}
\ee
where $V_n$ is the volume of the $n$-sphere and $'$ refers to the derivative
with respect to the $\delta$-function argument. This same charge
distribution arises in our present cases, too.  

    As an example, let us consider what happens if one of the
parameters, say $\ell_{n+1}$ is sent to zero.  It is clear from
(\ref{np1charge}) that the integration in (\ref{hintegral}) over the
associated direction $y_{n+1}'$ will become dominated by the
contribution from $y_{n+1}'$ close to zero, and so the
$(n+1)$-parameter charge distribution $\sigma_{n+1}$ in the
$\ell_{n+1}\longrightarrow 0$ limit will become the $n$-parameter
distribution
\be
\sigma_n(y_1,\ldots, y_{n+1}) = 
\delta(y_{n+1})\, \int_{-\infty}^\infty d\td y_{n+1}\,
\sigma_{n+1}(y_1,\ldots, y_n,\td y_{n+1})\,.
\ee
Evaluating the integral, we obtain
\be
\sigma_n =\fft{1}{2\, V_n\, \prod_{i=1}^{n} \ell_i}\,\Big( 2
(1-\sum_{i=1}^n \fft{y_i^2}{\ell_i^2})^{-1/2}\,  
\delta(1-\sum_{i=1}^n \fft{y_i^2}{\ell_i^2})
- (1-\sum_{i=1}^n \fft{y_i^2}{\ell_i^2})^{-3/2}\, 
\Theta(1-\sum_{i=1}^n \fft{y_i^2}{\ell_i^2})\Big)\, \delta(y_{n+1})\,.
\ee

   Sending another parameter, say $\ell_n$ to zero, we next obtain the
$(n-1)$-parameter charge distribution
\be 
\sigma_{n-1} = \fft{\pi}{V_n\, \prod_{i=1}^{n-1}\ell_i}\, \delta(1
-\sum_{i=1}^{n-1} \fft{y_i^2}{\ell_i^2})\, \delta^\2(y_{n},y_{n+1})\,.
\ee 
Further details of the successive results for smaller numbers $k$
of non-vanishing parameters $\ell_i$ are given in \cite{dist}.  Note that
the distributions associated with $k=n+1$ and $k=n$ non-vanishing
$\ell_i$ parameters both have regions with negative as well as
positive $p$-brane tensions.  For $k\le n-1$, on the other hand, the
distributions contain only positive tensions.

       When all the parameters $\ell_i$ vanish, corresponding to the
``vacuum'' domain-wall solution in $d=D-n$ dimensions, the
$D$-dimensional solution describes coincident $(D-n-2)$-branes at the
origin, with the constant 1 in the harmonic function $H$ dropped.
This can be viewed as a certain decoupling limit.  The metric of the
solution in the Einstein frame can then be expressed as
\be
ds_{\rm E}^2 = e^{\fft{a}{n-1}\phi}\, \Big(r^{n-3}\, dx^\mu dx_\mu +
\fft{dr^2}{r^2} + d\Omega_n^2 \Big)\,.
\ee
One can then define a dual frame $ds_{\rm dual}^2 = e^{-a
\phi/(n-1)}\, ds_{\rm E}^2$, in which the Lagrangian becomes
\be
{\cal L} = e\, e^{\fft{a(D-2)}{2(n-1)}\phi}\, \Big(
R  + \ft{(D-2)(n^2-n\, D - n + 3D -2)}{2(n-1)^2}\, 
 (\del\phi)^2 -\ft1{2n!}\, F_\n^2\Big)\,.
\ee
In this dual frame, the metric is AdS$\times S^n$ if $n\ne 3$, and
Minkowski$\times S^3$ when $n=3$.  This analysis was given in detail
in \cite{bst} for $D=10$, leading to the conjecture of a
Domain-wall/QFT correspondence.  Further studies of the
Domain-wall/QFT correspondence in general dimensions were given in
\cite{bh}.

         It is of interest to note that in the dual frame, the metric
depends only on the dimension $n$ of the internal sphere, but it is
independent of $D$; the $D$-dependence of the Einstein-frame metric
can all extracted as a conformal factor.  Note that the dual frame
metric has qualitative differences in the three situations $n>3$,
$n=3$ and $n<3$.  For $n=3$, the dual frame is Minkowskian, whilst for $n
\ne 3$, the spacetime is AdS.   However, for $n>3$ we have that $r=0$ is
the horizon, whilst for $n<3$ the horizon is instead at $r=\infty$.  These
qualitative differences have significance for the  structure of the spectrum
in the dual QFT, which we shall analyse in the next section.

    When the $\ell_i$ parameters are non-vanishing, the metric of the
distributed branes in the dual frame is given by
\be
ds_{\rm dual}^2 = \wtd \Delta^{\fft{n-3}{n-1}}\, ds_d^2 +
g^{-2}\, \wtd \Delta^{\fft{2}{n-1}}\, \sum H_i^2\, d\mu_i^2\,,
\ee
where
\be
ds_d^2 = (gr)^{n-3}\, dx^\mu dx_\mu + \fft{dr^2}{(gr)^2\, h^{\ft12}}
\,.\label{dualdw}
\ee
Again we see that the metric does not manifestly depend on $D$, but
on $n$ instead.

\section{Analysis of the spectrum}

   A minimally-coupled scalar field $\Phi$ obeys the wave equation
\be
\del_\mu(\sqrt{-g}\, g^{\mu\nu}\, \del_\nu \Phi)=0\,.
\ee
We make the Ansatz $\Phi = e^{ip\cdot x}\, \chi(r)$, where $m^2
=-p\cdot p$ determines the mass of the fluctuating mode, and so the
wave equation has the following general form
\be
r^{-1}\del_r\Big[r^{-1}\, \prod_{i=1}^{n+1}\sqrt{r^2 + \ell_i^2}
\,\del_r\chi\Big] = - {\cal Q}\,\chi\,,\label{genwave}
\ee
where ${\cal Q} = m^2\, g^{-\ft12(n+1)}$.   Remarkably,
the wave equation depends only on the dimension of the internal sphere, but
otherwise is independent of details of the original higher-dimensional
theory.

       It is helpful to cast the wave equation into the Schr\"odinger
form, which can be done by first writing the metric in a manifestly
conformally-flat frame as
\be
ds^2 = e^{2A(z)}\, (dx^\mu\, dx_\mu + dz^2)\,,
\ee
by means of an appropriate coordinate transformation.  The coordinate
$z$ runs from 0 to $z^*$, and $A(z)$ has the following
asymptotic behaviour:
\bea
e^{2A} \sim (z-z^*)^{\gamma^*}\,, \qquad \gamma^*=
-\fft{2(n+1)}{(d-2)(n-3)}& {\rm for}& z\rightarrow
z^*\,, \nn\\
e^{2A} \sim z^{\td\gamma}\,,\qquad
\td \gamma = \fft{2\gamma}{2-\gamma}=-\fft{2(n+1-k)}{(d-2)(n-3-k)}
\,,
&{\rm for} & z\rightarrow 0\,.\label{conformal}
\eea

   Making the field redefinition $\chi = e^{-(D-2)A/2}\, \psi$, the
wave equation assumes the form
\be
(-\del^2 - V)\, \psi = \ft14 {\cal Q}\, \psi\,,
\ee
with the Schr\"odinger potential given by
\be
V = \fft{d-2}{2}\,  A'' + \fft{(d-2)^2}{4} \, (A')^2\,.
\ee
The asymptotic behavior of the potential is given by
\bea
V&\sim& \fft{c^*}{(z-z^*)^2}\,,\qquad \hbox{for}
\qquad z\rightarrow z^*\,,\nn\\
V&\sim& \fft{c}{(z-{\tilde z})^2}\,,\qquad \hbox{for}\qquad z\rightarrow
{\tilde z}
\,,\label{potasym}
\eea
where
\be
c^*=-\ft14 + \fft{(n-1)^2}{(n-3)^2}\,,\qquad
c=-\ft14 + \fft{(n-1-k)^2}{(n-3-k)^2} \ge -\ft14\,.
\label{cvalue}
\ee
The range of the coordinate $z$ is determined by the values of $z^*$
and $\tilde z$, which in the original coordinate $r$ correspond to
$r\to \infty$ and $r\to 0$ limit, respectively. It is understood that
if $z^*$ or $\tilde z$ equals $\pm \infty$, the potential in
(\ref{potasym}) is of the form $\pm 1/z^2$.

Note that for $n\le 3$ [$n\ge 4$] the limit $r\to\infty$ corresponds
to $z\to z^*$ with $z^*=\infty$ [$z^*={\rm finite}$].  On the other
hand for $n-k\le 3$ [$n-k\ge 4$] the limit $r\to 0$ corresponds to
$z\to {\tilde z}$ where ${\tilde z}=0$ [${\tilde z}= -\infty$].  
When $n=3$ or
$k=n-3$, where the denominator of the above expression vanishes, the
coordinate $z$ depends logarithmically on the original coordinate $r$
($z\sim \log(r)$) and the Schr\"odinger potential becomes constant:
$V=1/4$.

      Note that since the wave equation is independent of $D$, whilst
the metric depends on $D$, it may be more instructive to perform a field
redefinition directly on the wave equation (\ref{genwave}).  This can
be done by first defining $y=r^2$, and then introducing a new coordinate
$z$ defined by $\del y/\del z = \sqrt{f(y)}$, 
where $f(y)=[\prod_{i=1}^{n+1}(y+\ell_i^2)]^{1/2}$. (These are 
 the defining equations that relate $z$ and $r$ coordinates.)  
 The Schr\"odinger potential is then given by
\cite{dist}
\be
V=\ft14 \del_z^2 \log f + \ft1{16} (\del_z\log f)^2\,.
\ee
and it clearly depends on $n$ and $\ell_i$  ($i=1,\cdots , k$) only.

\subsection{Vacuum excitations}

           When all the parameters $\ell_i$ vanish, the solution
(\ref{gendw}) becomes a domain-wall vacuum solution.  In the case when
$a^2=0$, which occurs for $(D,n)=(11,7)$, $(10,5)$ and $(11,4)$, the
solution is just the AdS spacetime in $d=4$, 5 and 7 respectively.  For
$a^2\ne 0$, the metric of the solution is (\ref{vacuum}).  The
metric is flat near $\rho=\infty$, but becomes singular as $\rho$ approaches
zero.  Since we have
\be
\fft{4(n+1)}{a^2(D-2)} - 2= \fft{4(d-2)(n-3)}{a^2(D-2)}\,,
\ee
the singularity is marginal for $n\ge 3$, but naked for $n < 3$.

        The characteristics of the Schr\"odinger potential depend
only the value of $n$.  For $n=0,1,2$, the potential is given by
\be
V=\fft{c^*}{z^2}\,,\label{cvaluevacuum}
\ee
where $c^*$ is given in (\ref{cvalue}).  The coordinate $z$ runs from
0 to infinity as $r$ runs from 0 to infinity.  For $n=3$, the
potential is a constant, $V=1/4$, and the coordinate $z$ runs from minus
infinity to infinity as $r$ runs from 0 to infinity.  For $n\ge 4$,
the potential is of the same form as (\ref{cvaluevacuum}), but the
coordinate $z$ now runs from minus infinity to 0 as $r$ runs from 0 to
infinity.  Thus we see that although the domain-wall vacuum can have
(naked) singularities, the quantum fluctuations are nevertheless well
behaved.  In fact it is straightforward to solve the minimally-coupled
scalar wave equation in the domain-wall vacuum, namely
\be
r^{-1}\del_r (r^n\, \del_r\chi) = -{\cal Q}\, \chi\,.\label{chieq}
\ee
If we define a new dependent variable $y$ by
\be
\chi(r) = y(r)\, r^{-(n-1)/2}\,,
\ee
and change to the new independent variable $z$ defined by
\be
z = \fft{2\sqrt {\cal Q}}{n-3}\, r^{-(n-3)/2}\,,
\ee
then the wave equation (\ref{chieq}) becomes the Bessel
equation
\be
z^2\, y''(z) + z\, y'(z) + (z^2-\nu^2)\, y(z)=0\,,
\ee
where
\be
\nu = \fft{n-1}{n-3}\,.
\ee
The solutions to (\ref{chieq}) are therefore given by
\be
\chi(r) =a\,  
r^{-(n-1)/2}\, J_\nu\Big(\fft{2\sqrt{\cal Q}}{n-3}\, r^{-(n-3)/2}\Big)
+b\, 
r^{-(n-1)/2}\, Y_\nu\Big(\fft{2\sqrt{\cal Q}}{n-3}\, r^{-(n-3)/2}\Big)
\,.
\ee
A special case arises for $n=3$ (the Schr\"odinger potential is constant,
$V=1/4$,
there) for which we find
\be
\chi(r)= a \, r^{-1+\im\, \sqrt{{\cal Q} -1}} + 
               b\, r^{-1-\im\, \sqrt{{\cal Q} -1}}\,.\label{n3vac}
\ee
The requirement that $Q\ge 1$ corresponds to the condition that there is an
energy gap.

\subsection{Domain-wall excitations}

         When some of the $\ell_i$ parameters are non-vanishing, the
wave equations cannot in general be solved explicitly.  Here, for
simplicity, we shall consider the case where all the non-vanishing
$\ell_i$ are equal.  There are certain examples where the wave
equations can be solved exactly.  Two of these ($n=5$, $k=2$ with two
equal charges $\ell_i$, and $n=5$, $k=4$ with four equal charges) are
solved in \cite{FGPW}.  Another solvable example is $n=3$, $k=2$, with
the two non-vanishing charges equal, say $\ell_1=\ell_2\equiv\ell$.
In this case, if we let $x=-r^2/\ell^2$, equation (\ref{genwave})
becomes the hypergeometric equation
\be
x(1-x)\, \chi'' + (1-2x)\, \chi' - \ft14 {\cal Q}\, \chi=0\,,\label{hypereq}
\ee
and so one solution gives
\be
\chi_1 = {}_2F_1[a,b;1;-\fft{r^2}{\ell^2}]\,,\qquad a=
\ft12+\ft{\im}{2}\, \sqrt{{\cal Q}-1}\,,\qquad 
b=\ft12-\ft{\im}{2}\, \sqrt{{\cal Q}-1}\,.\label{hypergeom}
\ee
Note again that $Q>1$ corresponds to the condition that the
(continuous) spectrum has a gap owing to the properties of the
Schr\"odinger potential $V\ge 1/4$ (figure g).  Note that at small $r$ we
therefore have $\chi_1\sim 1$, while at large $r$ the asymptotic
behaviour is of the same form as in (\ref{n3vac}).  Since the $c$
argument of the hypergeometric function $_2F_1[a,b;c;x]$ in
(\ref{hypergeom}) is an integer, the second solution $\chi_2$ of
(\ref{hypereq}) must be obtained by taking an appropriate limit of the
standard second solution $x^{1-c}\, _2F_1[a-c+1,b-c+1;2-c;x]$.  This
gives a logarithmic behaviour of the form $\chi_2\sim \log r$ at small
$r$.

    For the remaining examples, although we cannot solve the wave
equation analytically we can determine the structure of the spectra 
for the various cases from the forms of their Schr\"odinger potentials.
The results are summarised in Table 1.

\bigskip\bigskip

\centerline{
\begin{tabular}{|c|c|c|c|c|} \hline
$n$  & $k$ & $z$-range & $V$ type & Spectrum \\ \hline
0,1  & 0,1 & $(0, \infty)$ & a & continuous \\ \hline
2    & 0   &  $(0, \infty)$ & b & continuous \\ \hline
     & 1,2 &  $(0, \infty)$ & c & continuous \\ \hline
3    & 0   & $(-\infty, \infty)$ & $V=\ft14$ & cont. with gap \\ \hline
     & 1   & $(0, \infty)$  & d & disc., cont. with gap \\ \hline
     &2,3  & $(0, \infty)$ & e & cont. with gap \\ \hline\hline
$\ge 4$ & $\le n-4$ & $(-\infty,0)$ & f & continuous \\ \hline
        & $n-2$ & $(-\infty,0)$ & g & cont. with gap \\ \hline
        & $n-3$& $(-1,0)$ & h & discrete \\ \hline
        &$n,n-1$ & $(-1,0)$ & i & discrete \\ \hline
\end{tabular}}

\bigskip

\centerline{Table 1: Spectral analysis for domain-wall solutions for 
various $n$'s and $k$'s.}

\bigskip

     The various different types of structures of the potentials are
sketched in Figure 1.  

\centerline{
\epsfig{file=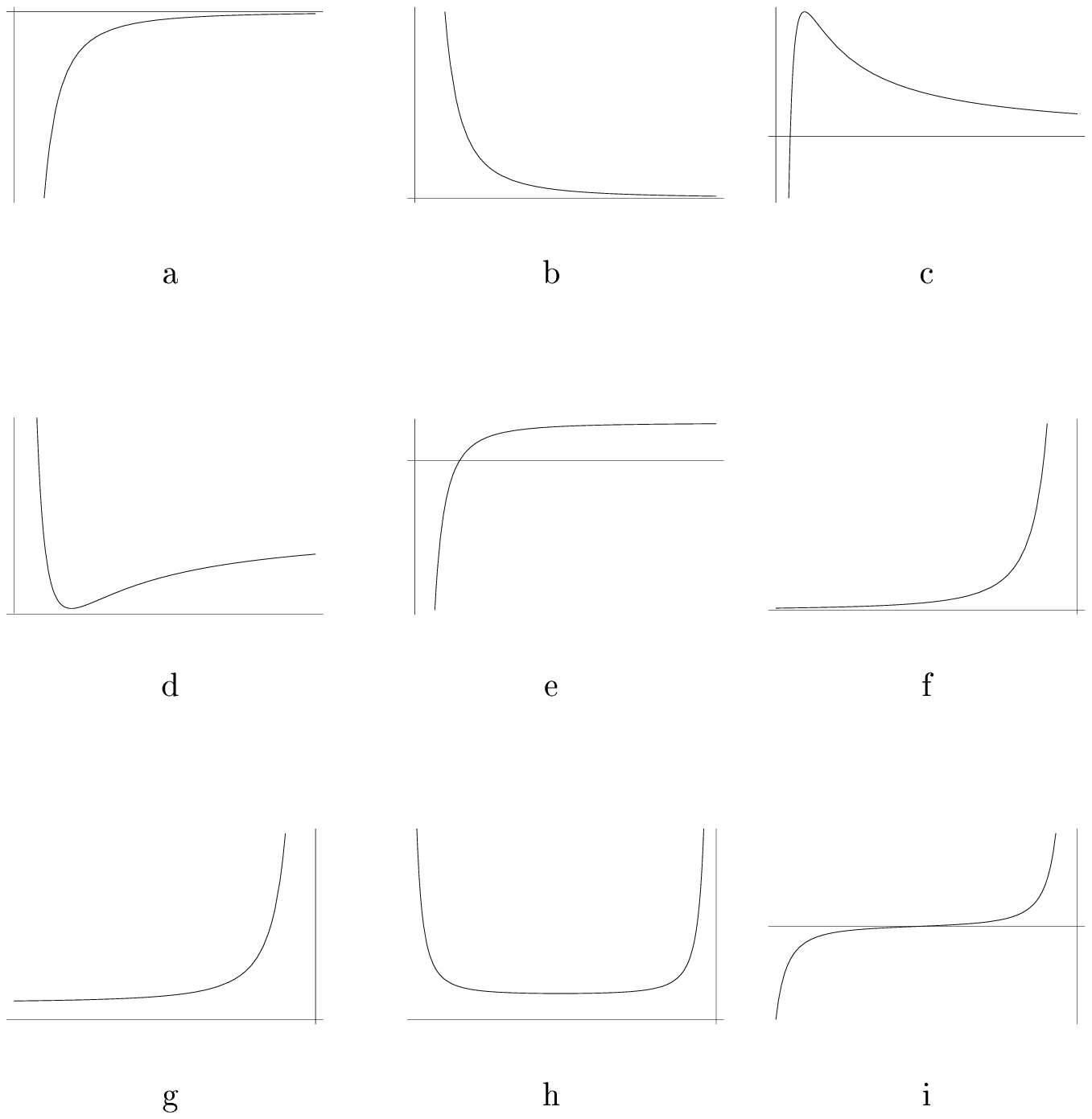,width=15cm}
}

\vspace{-6cm}
\centerline{Figure 1: Sketches of the various Schr\"odinger potentials}

\section{Conclusions}

         In this paper we have studied consistent $n$-sphere
reductions of a $D$-dimensional theory of gravity coupled to an
$n$-form field-strength and a dilaton. Provided that the dilaton has a
specific strength of coupling to the $n$-form, given by
(\ref{avalue}), we have proven the
consistency of the non-linear Kaluza-Klein Ansatz for the $n$-sphere
reduction in which there are $n$ scalars parameterising
right-ellipsoidal inhomogeneous deformations of the
sphere.\footnote{We did not turn on the Kaluza-Klein gauge-fields in
the reduction, which corresponds
to a consistent truncation of the theory.  However in
the appendix we also discuss an $n$-sphere reduction of this
Lagrangian that corresponds to making pair-wise identifications of the
diagonal scalar fields, together with turning on the electric
components of the Abelian Kaluza-Klein fields.  This reduction
provides a $D$-dimensional embedding of the $(D-n)$-dimensional
non-extreme (large) charged-black holes as (near-extreme) spinning
electric $(D-n-2)$-branes. In the BPS limit the charged black holes
become neutral BPS domain-wall solutions.}

      The generality of these consistent reductions provides a
framework within which we can address the $D$-dimensional embedding
of a class of solutions of the reduced gauged supergravities in 
$d=D-n$ dimensions. In general, these gauged supergravities
have potentials for the scalar fields that do not admit
AdS ground-states, and thus in general, the typical solutions
correspond to domain walls that are asymptotic to the ``dilatonic''
vacuum. In particular, we found the general class of BPS
domain-wall solutions that are specified by $k=\{1,\cdots ,n\}$
parameters, which characterise the harmonic functions of $k$ non-trivial
scalars.

    All these solutions  have explicit representations as continuous
distributions of extremal $(D-n-2)$-branes, and thus in the context
of  the Domain-wall/QFT correspondence they describe the Coulomb
phase of the  dual strongly-coupled field theory.

   The universal properties of these gravity solutions manifest
themselves in the properties of the wave equations in these
backgrounds.  For minimally-coupled scalars, the wave equations are
completely universal and depend only on the dimension $n$ of the
compactifying sphere and the number $k$ of parameters in the harmonic
functions specifying the non-trivial scalar fields.  Remarkably, the
wave equations are {\it independent} of the original dimension
$D$. Thus in the dual field theory the bound-state spectrum is
completely specified by $n$ and $k$. We gave an analysis of the
spectra for all these cases.

   One of the interesting outcomes of our study is the generality and
universality of the BPS solutions for the specific subsector of the
sphere-reduced gravity theories.  This provides a strong indication
that the dual field theories should exhibit the same intriguing
features, irrespective of the dimension.

\section*{Acknowledgments} 

       We should like to thank K. Skenderis for a useful
discussion. C.N.P. would like to thank High Energy Theory Group at
Penn for hospitality.

\appendix

\section{Single-charge rotating $p$-branes}

    The Lagrangian (\ref{genlag}) also admits rotating $p$-brane
solutions.  In this appendix, we show that such a rotating $p$-brane
associated with $a$ given by (\ref{avalue}) can be dimensionally
reduced on the transverse spherical space, and it then gives rise to a
domain-wall black hole in the lower dimension.  The Lagrangian
(\ref{genlag}) admits an electric $(d-1)$-brane with $d=n-1$, or a
magnetic $(d-1)$-brane with $d=D-n-1$.  We shall consider only the
magnetic solution here, since the electric one can be viewed as a
magnetic solution of the dual $(D-n)$-form field strength
$F_{\sst{(D-n)}}$.  There are two cases arising, depending on whether
$\td d $ is even or odd.

\bigskip\bigskip
\noindent{\underline{{\it Case 1:}\ \ $n= 2N-1$}}
\bigskip

       In this case, there are $N$ angular momenta $\ell_i$, with
$i=1,2,\ldots, N$.  We find that the metric of the rotating
$(n-2)$-brane solution to the equations following from (\ref{genlag})
is \cite{ten}
\bea
ds_{D}^2&=& H^{-\ft{n-1}{D-2}}\Big(-(1 - \fft{2m}{r^{n-1}\Delta})\,
dt^2 + d\vec x\cdot d\vec x\Big) +
H^{\ft{D-n-1}{D-2}}\Big[\fft{\Delta\, dr^2}{H_1\cdots H_N -
\fft{2m}{r^{-(n-1)}} }\nn\\
&&+r^2 \sum_{i=1}^N H_i(d\td\mu_i^2 + \td\mu_i^2\, d\phi_i^2) -
\fft{4m\cosh\a}{r^{n-1}\, H\, \Delta}\, dt\,
(\sum_{i=1}^N \ell_i\, \td\mu_i^2\, d\phi_i)\nn\\
&&+\fft{2m}{r^{n-1}\, H\, \Delta}\,
(\sum_{i=1}^N \ell_i\, \td\mu_i^2\, d\phi_i)^2 \Big]\ ,
\label{tddeven}
\eea
where the functions $\Delta$, $H$ and $H_i$ are given by
\bea
&&\Delta = H_1\cdots H_N\, \sum_{i=1}^N \fft{\td\mu_i^2}{H_i}
\ ,\qquad H= 1 + \fft{2m\, \sinh^2\a}{r^{n-1}\, \Delta}\ ,
\nn\\
&&H_i = 1 + \fft{\ell_i^2}{r^2}\ ,\qquad i=1,2,\ldots, N\ .
\label{variousdef}
\eea
The dilaton $\phi$ and the field strength $F_\n$ are given by
\be
e^{2\phi/a} = H\ ,\qquad
e^{a\phi}{*F_\n}= \fft{dH^{-1}}{\sinh\a}\wedge\Big(\cosh\a \, dt +
\sum_{i=1}^N \ell_i\, \mu_i^2\, d\phi_i\Big )\wedge d^{D-n-2}x\ .
\ee
The $N$ quantities $\td\mu_i$ are subject to the constraint
$\sum_i \td\mu_i^2=1$.  They are related to our previous coordinates
constrained $\mu_i$ on the sphere as follows:
\be
\mu_1 + \im\, \mu_2 = \td\mu_1\, e^{\im\,\phi_1}\,,\qquad
\mu_3 + \im\, \mu_4 = \td\mu_2\, e^{\im\,\phi_2}\,,\qquad \hbox{\it etc}.
\label{muredef}
\ee
     
    We now consider the decoupling limit, which is obtained by
making the rescalings
\bea
&&m\rightarrow \epsilon^{n-1}\, m\,,\qquad \sinh\a \rightarrow
\epsilon^{-(n-1)/2}\, \sinh\a\,,\nn\\
&&r\rightarrow \epsilon\, r\,,\qquad x^\mu\rightarrow \epsilon^{-2}\, 
x^\mu\,,\qquad \ell_i\rightarrow \epsilon\,\ell_i\
\eea
and then sending $\epsilon\rightarrow 0$.  In this limit, the additive
constant 1 in the function $H$ in (\ref{variousdef}) can be dropped.
Furthermore, the last term in (\ref{tddeven}) can also be dropped.
The remaining metric can be expressed as
\be
ds_D^2 = Y^{\fft{2}{D-2}}\Big(\Delta^{\fft{n-1}{D-2}}\, ds_{d}^2 +
g^{-2}\,\Delta^{-\fft{(D-n-1)}{(D-2)}}\sum_{i=1}^{N} \bar X_i^{-1} 
(d\td\mu_i^2 + \td\mu_i^2(d\phi_i + g\, A_\1^i)^2)\Big)\,,\label{u1ansatz}
\ee
where $\Delta = \sum \bar X_i\, \td\mu_i^2$, and $g=(2m\,
\sinh^2\a)^{-1/(n-1)}$.   The $d=D-n$ dimensional metric and
the scalar fields $X_i$ are given by
\bea
ds_{d}^2 &=& -h^{-\ft{d-3}{d-2}}\, f\, dt^2 + h^{\ft{1}{d-2}}
\Big(\fft{dr^2}{(gr)^{5-n}\, f} + (gr)^{n-3}\, d\vec x\cdot d\vec x
\Big)\,,\nn\\
X_i&=&(gr)^{\ft{a^2\,(D-2)}{4(d-2)}}\, h^{\ft{(d-3)}{2(d-2)}}\, H_i^{-1}
\,,\nn\\
Y&=&\prod_{i=1}^N X_i = \Big((gr)^{n+1}\, 
h^{-1}\Big)^{\ft{a^2\, (D-2)}{8(d-2)}}\,,\nn\\
A_\1^i&=& \fft{1-H_i^{-1}}{g\,\ell_i\, \sinh\a}\, dt\,,\qquad
h=\prod_{i=1}^N H_i\,.\nn\\
f&=& (gr)^{n-3} (h - \fft{2m}{r^{n-1}})\,.
\eea
This solution describes $N$ electrically-charged black holes in a
$d$-dimensional domain-wall background.

   Note that in general, the abstract metric Ansatz that we have
written in (\ref{u1ansatz}) does {\it not} (at least as it stands)
correspond to part of a consistent Kaluza-Klein reduction.  It can be
viewed as a modification of the general consistent metric Ansatz in
(\ref{ansatz}) in which we first (consistently) partially truncate the
scalars by setting them equal in pairs ($X_1=X_2=\bar X_1$,
$X_3=X_4=\bar X_2$, {\it etc}\.).  Then, having also made the
redefinitions (\ref{muredef}), we introduce a $U(1)$ gauge field
$A_\1^i$ associated with the rotation in each of the original 2-planes
$(\mu_1,\mu_2)$, $(\mu_3,\mu_4)$, {\it etc}.  (Although we have not
presented it here, we can also straightforwardly carry out the same steps
on the original Ans\"atze for $\hat\phi$ and $\hat F_\n$ in
(\ref{ansatz}) too.)  This {\it does} give a
consistent reduction in the case $(10,5)$ discussed in \cite{ten}, but
in general additional fields would have to be included too.  The
reason for this is that the $U(1)$ gauge fields, in quadratic products
of the form $F^i_\2\wedge F^j_\2$, will act as sources for other
fields.  In the special case of $(D,n)=(10,5)$, they actually act as
sources for themselves (corresponding to cubic Chern-Simons terms in
the five-dimensional theory), but in the other cases they will act as
sources for additional fields, requiring a larger set of fields in the
Kaluza-Klein reduction Ansatz.

    The metric (\ref{u1ansatz}), together with analogously-obtained
expressions for $\hat\phi$ and $\hat F_\n$, is nevertheless still
usable in appropriate circumstances.  The problematic terms
$F^i_\2\wedge F^j_\2$ actually vanish for our specific domain-wall
black hole solutions since all the $U(1)$ charges are purely electric.
This means that these particular lower-dimensional configurations will
lift to the higher dimension without necessitating the turning-on of
the additional fields that would be needed for a fully-consistent
Ansatz, but which have been omitted in our discussion.  Thus we still
have an exact embedding of these specific solutions in the higher
dimension.

\bigskip\bigskip
\noindent{\underline{{\it Case 2:}\ \ $n = 2N$}}
\bigskip

   Here, the solution has the same form as (\ref{tddeven}), but with
the range of the index $i$ extended to include 0.  However, there is
no angular momentum parameter or azimuthal coordinate associated with
the extra index value, and so $\ell_0=0$ and $\phi_0=0$.  The
$\td\mu_i$ and $\phi_i$ coordinates are now related to the original
coordinates $\mu_i$ on the sphere by
\be
\mu_0=\td\mu_0\,,\qquad
\mu_1 + \im\, \mu_2 = \td\mu_1\, e^{\im\,\phi_1}\,,\qquad
\mu_3 + \im\, \mu_4 = \td\mu_2\, e^{\im\,\phi_2}\,,\qquad \hbox{\it etc}.
\label{muredef2}
\ee
Otherwise,
all the formulae in Case 1 generalise to this case, simply by extending
the summation to span the range $0\le i\le N$.  Of course $H_0=1$ as a
consequence of $\ell_0=0$.

       Note that for $a=0$, we have $(D,n)=(11,7), (11,4)$ and $(10,5)$.
These correspond to the rotating M-branes \cite{rms,ten} and D3-branes
\cite{KLT,rd32,ten}.

\end{document}